\documentclass[12pt,letterpaper]{article}        % For preprint
\pdfoutput=1
\usepackage[usenames,dvipsnames]{color}
\usepackage{jheppub}

\newcommand{\Daniele}[1]{{\color{Blue} #1}}

\title{\Large Chasing the cuprates with dilatonic dyons}
\author[a]{Andrea Amoretti,}
\author[b]{Matteo Baggioli,}
\author[c]{Nicodemo Magnoli,}
\author[d]{Daniele Musso,}

\affiliation[a]{Department  of  Applied  Mathematics  and  Theoretical  Physics, University  of  Cambridge, Cambridge, CB3 OWA, UK}

\affiliation[b]{Institut de F\'{i}sica d'Altes Energies (IFAE), Universitat Aut\`{o}noma de Barcelona, The Barcelona Institute of Science and Technology, Campus UAB,\\ 08193 Bellaterra (Barcelona), Spain }

\affiliation[c]{ Dipartimento di Fisica, Universit\`a di Genova,\\
via Dodecaneso 33, I-16146, Genova, Italy\\and I.N.F.N. - Sezione di Genova}

\affiliation[d]{Abdus Salam International Centre for Theoretical Physics (ICTP)\\
Strada Costiera 11, I-34151 Trieste, Italy}

\emailAdd{aa917@damtp.cam.ac.uk} \emailAdd{mbaggioli@ifae.es} \emailAdd{nicodemo.magnoli@ge.infn.it} \emailAdd{dmusso@ictp.it}

\abstract{Magnetic field and momentum dissipation are key ingredients in describing 
condensed matter systems. We include them in gauge/gravity and 
systematically explore the bottom-up panorama of holographic IR effective field theories
based on bulk Einstein-Maxwell Lagrangians plus scalars.
The class of solutions here examined
appear insufficient to capture the phenomenology of charge transport in the cuprates.
We analyze in particular the temperature scaling of the resistivity and of the Hall angle.
Keeping an open attitude, we illustrate weak and strong points of the approach.}

\preprint{DAMTP-2016-23}

\begin{document}

\maketitle
\pagestyle{plain} \setcounter{page}{1}
\newcounter{bean}
\baselineskip16pt

\section{Introduction and motivation}

The ubiquitous presence of a magnetic field in experimental setups and - more importantly - 
its direct r\^ole in investigating the strange metal criticality calls for a solid theoretical account.
We refer in particular to the recent experimental proposal highlighting that the strange metal physics
(and specifically the transport response) could be fruitfully studied focusing on the competition 
of the scales dictated by the magnetic field and the temperature \cite{Hayes}. One of the main results of our analysis 
consists in underlying and specifying explicitly the importance of the magnetic field $B$ in affecting equilibrium and transport 
properties of a strongly coupled field theory with momentum dissipation modeled through gauge/gravity techniques.

The holographic framework we adopt relies upon conjectures and various approximations on which we need to meditate 
carefully (later sections are dedicated to comments on the delicate points). The aim at stake is to build effective field theories capable of describing the 
low-temperature behavior of condensed matter systems with particular attention to the scaling properties
in $T$ and corrections in $B$ of the various thermodynamic and linear response quantities%
\footnote{This idea is part of a long-standing program within the holographic community \cite{Charmousis:2010zz,Gouteraux:2011ce,Gouteraux:2011xr}.}. 
The attitude is in general phenomenological
or bottom-up. Namely we choose gravity models that in a gauge/gravity spirit are conjectured to be dual 
to strongly coupled field theories. More precisely, we resort to a wide class of bulk models described by Einstein-Maxwell 
Lagrangians plus scalars; they feature the ``axion-like'' scalars accounting for momentum dissipation
(and realizing a massive gravity through a Stueckelberg mechanism%
\footnote{See \cite{Baggioli:2014roa,Alberte:2015isw} for more general holographic models making use of the Stueckelberg scalars 
and their connection with theories of massive gravity and Condensed Matter phenomenology. 
It is relevant to refer also to the wider non-holographic context where similar ideas have 
been proposed and explored, see for instance \cite{Dubovsky:2011sj,Nicolis:2015sra}.})
and a ``dilaton-like'' scalar encoding a non-trivial renormalization group flow. 

We focus on infrared solutions only. This means that the various phenomenological functions 
of the model - specifically, the kinetic functions for the gauge field and momentum-dissipation scalars
as well as the potential for the dilaton-like field - are approximated by means of a single exponential term
which is the dominant contribution in the infrared. The input parameters into the game are therefore represented 
by the exponents of these exponential terms alongside the parameters coming from the dilaton black-brane solutions.
These latter are in general of the hyperscaling-violating and Lifshitz type, characterized by three scaling exponents.
Two of them, $\theta$ and $z$, appear directly in the metric and are associated to the ``effective dimension'' of the theory and the different scaling of space and time, respectively%
\footnote{See also \cite{Goldstein:2009cv,Charmousis:2010zz,Goldstein:2010aw,Gouteraux:2011ce,Hartnoll:2009ns,Donos:2014oha,Kiritsis:2015doa,Kiritsis:2015yna,Kiritsis:2015oxa,Alishahiha:2012qu,Fadafan:2012hr,Hoyos:2010at,Dong:2012se}
for previous analyses about transport properties of holographic systems with Lifshitz and hyperscaling violating geometries.}. 
The third one, $\zeta$, parameterizes the IR anomalous dimension for the charge density operator and it affects directly the scaling of the conductivity%
\footnote{One direct way of producing an anomalous dimension for the gauge field is provided by the unparticles 
models \cite{Karch:2015zqd}. See also \cite{Karch:2015pha} for an alternative approach.} 
(see \cite{Gouteraux:2013oca,Gouteraux:2012yr,Gouteraux:2014hca,Hartnoll:2015sea,Karch:2014mba}).\color{black}

Such a class of models is at the core of a long-standing program to apply holography to condensed matter systems
and constitutes a natural setup to describe IR criticality associated to a scaling fixed point.
It provides a flexible and general framework for a phenomenological description of possible 
gauge/gravity models aiming to exhaustive classifications; notably, a wide subclass of such 
models features a vanishing residual entropy (further comments on this will prove to be in order).
Moreover, some top-down glimpses have encouraged in convincing of the general 
relevance of this IR setup (see for instance \cite{Gubser:2009qt}). Roughly, the philosophy consists in taking the minimal set of ingredients and allowing 
for the most general behavior which does not spoil completely the infrared geometry. All the actors on stage -
the chemical potential, the magnetic field and the momentum dissipation device - are allowed to be either marginally relevant
(roughly, weighting as much as the dilaton-like scalar in the IR flow) or irrelevant. Specifically,
we start from an analytical background solution encoding the marginally relevant quantities 
and then deform it with perturbative irrelevant operators%
\footnote{The method is clarified in the following sections and builds on the techniques described in \cite{Kundu:2012jn}.
The possibility of perturbing by means of relevant operators will be also commented later.}.

We focus on bulk solutions that are nearly extremal, obtained perturbing slightly their extremal counterparts.
Both the temperature and the irrelevant operators are therefore perturbations; focusing on the entropy
we show that it is possible to consider both hierarchies and take either $T$ or $B^2$ as a small parameter 
when compared to the other scales at play. This argument is based only on dimensional analysis and scaling properties 
and similar considerations hold for the transport quantities as well.

On the phenomenological side, it is important to recall that 
experimental evidence \cite{Cooper603} shows that the characteristic cuprate scaling
features, namely 
\begin{equation}
\label{cupintro}
 \rho_{xx}\propto T\qquad \text{and}\qquad \cot \theta_H \propto T^2 \ ,
\end{equation}
keep being present and robust also in the regime of strong magnetic field $B$.
It is important to keep in mind however that an arbitrarily low-$T$ regime requires some extra caution about
the reliability of the models%
\footnote{Let us stress that also the experimental results have
some low-temperature cut-off; the study of EMD geometries at exactly zero temperature
could then be of no immediate phenomenological relevance or, at least, it does not constitute a primary worry.}, 
and specifically the supergravity approximation; 
we discuss these subtleties and the difficulties of providing sharp bounds on $T$ \emph{a priori} of a clear 
string completion. 
At the outset it is important to underline that, as long as the charge density is a marginal or 
irrelevant operator (which are the two cases considered in detail in the present article), the cuprates 
scalings \eqref{cupintro} can not be attained. We clarify later this point and comment on possible extensions 
to cases with a relevant charge density operator.

\section{Results}

In this paper we mainly consider the cases in which the magnetic field constitutes an irrelevant perturbation
and comment on other possibilities.
In the models at hand we find that, asking for 
the magnetic field to be irrelevant, automatically implies that the effective dimension $d_{\theta}=d-\theta$ and the 
Lifshitz exponent $z$ are negative. Even though this fact is strange when one thinks at standard effective 
quantum field theories, we do not find any ``in principle'' reason to exclude these solutions on the gravitational side%
\footnote{In this direction it would be interesting to perform an analysis along the lines of \cite{Hoyos:2010at}
in the presence of hyperscaling violation. One could also argue that $\theta<0$ could be no worrisome on the basis 
that there exist Condensed Matter models featuring \emph{dimensional crossover} \cite{dimcross}.}. 
Consequently, we consider these black hole solutions as holographic duals of exotic strongly coupled quantum 
field theories keeping in mind that we do not know any realization of an effective standard quantum field theory 
where the effective dimension and the Lifshitz exponent are negative%
\footnote{The same holds for a marginal $B$, see the dyonic solution of \cite{Gath:2012pg}
which leads to $\theta=4$.}.

One important result of the present analysis is that, also in the presence of a magnetic field, the picture 
outlined in \cite{Donos:2014uba,Gouteraux:2014hca} remains valid: namely, as long as the charge density is marginal, the momentum-dissipation part 
of the conductivity always dominates the transport (in contrast with the basic requirement of \cite{Blake:2014yla})%
\footnote{We refer to the inverse Matthiessen rule (at $B=0$)
\begin{equation}
 \sigma = \sigma_{0} + \sigma_{\text{diss}}\ ,
\end{equation}
(see for instance \cite{Blake:2013bqa}). The $\sigma_{0}$ term refers to the DC contribution at zero net heat current 
(which, at finite density, is in general different from the contribution at zero momentum current \cite{Davison:2015bea}); 
$\sigma_0$ adds up with a contribution $\sigma_{\text{diss}}$ 
proportional to the charge density and it is directly connected to the momentum-dissipating dynamics.
Note however that this splitting of the conductivity in two terms can be quite tricky in generic holographic effective theories. 
When one couples directly the charged sector with the momentum-dissipating sector, or 
in the case of a non-linear extension of the Maxwell action, the first term of the conductivity $\sigma_0$ 
can depend on the charge density and the momentum-dissipation rate as well 
(see e.g. \cite{Davison:2015bea,Blake:2015hxa,Baggioli:2016oqk,Baggioli:2016oju}). 
In order to identify properly the various time scales in the system one can resort to an analysis
of the QNM spectrum because the DC transport features are in general not sufficient in providing such 
information \cite{Davison:2015bea}.}
and the Hall angle and the resistivity have the following behavior:
\begin{equation}
\label{pred1}
\cot \theta_H \propto T^n \ , \qquad \rho \propto T^{n} \ , 
\end{equation}  
which evidently makes it impossible to fit the strange metals picture. 

In the scaling geometries analyzed with the techniques adopted in this paper, the only way to achieve a ``non-Fermi liquid like'' 
behavior for the transport coefficients, namely different scalings for $\cot \theta_H$ and $\rho$, is to consider 
the charge density as an irrelevant perturbation. In such a case, the conductivity at zero net heat current 
dominates the momentum-dissipation piece leading to different scalings for the Hall angle and the resistivity, namely:
\begin{equation}
\label{pred2}
\cot \theta_H \propto T^n \ , \qquad \rho  \propto T^m \ .
\end{equation}
However, also in this case, the irrelevance of the charge density operator implies that $m>n$, which excludes the cuprates scenario%
\footnote{The scalings \eqref{pred1} and \eqref{pred2} can be predicted from the scalings worked out by \cite{Karch:2014mba}, keeping in mind
that $\Phi=\theta-d$ for EMD solutions with a marginal charge density in
the IR, while $\Phi \ne \theta - d$ otherwise (as shown in \cite{Gouteraux:2014hca,Hartnoll:2015sea}).}. 

In Appendix $C$ we extend the analysis to the Seebeck coefficient accounting for the thermopower.
We do not consider the magneto-resistance because it is quadratic in $B$ and this coincides with the 
perturbative order in $B$ that we neglect.

Finally, concerning the case in which the magnetic field is a relevant perturbation, a detailed numerical 
analysis would be needed%
\footnote{For instance an explicit computation is required to shed conclusive light to the 
interplay of the various scales and the radial region where the transition from the intermediate scaling geometry to 
the ultimate IR point occurs.}.
However, one can think in analogy with the analysis of \cite{Kundu:2012jn} and consider that,
as long as the momentum-dissipation operator is marginal or irrelevant, the nature of the IR 
fixed point found in \cite{Kundu:2012jn} should not be modified. Then one can argue that, also in this case, the 
cuprate picture cannot be found. In fact, according to \cite{Kundu:2012jn} the IR fixed point should be of 
the Reissner-Nordstr\"{o}m type. This kind 
of black hole solutions do not presents the correct scalings for the Hall angle and the resistivity \cite{Amoretti:2015gna}.

\section{Model}
\label{modello}

The gravity model we consider features an Einstein-Maxwell action supplemented by ``axion-like'' scalars
whose role is to incorporate momentum dissipation \cite{Andrade:2013gsa} and a ``dilaton-like'' field possessing a non-trivial potential and appearing 
in the kinetic functions of the other matter fields. All the fields are electromagnetically neutral. The explicit form of the action is
\begin{equation}\label{model}
\mathcal{S}=\int d^4x\, \sqrt{-g} \left[R-2\left(\nabla \phi \right)^2-f(\phi)F_{\mu \nu}F^{\mu \nu}-V(\phi)-\frac{1}{2}Y(\phi) \sum_{I=1}^2\left(\nabla \psi^I\right)^2 \right]\ .
\end{equation}
The functions $f(\phi)$, $V(\phi)$ and $Y(\phi)$ are ``phenomenological'', namely not derived in a top-down sense; the same being true
for the field content. Moreover they are assumed to take the following form
\begin{equation}\label{pheno}
f(\phi)=e^{2\alpha \phi} \, \qquad V(\phi)=-\left|V_0\right| e^{2\delta \phi} \ , \qquad Y(\phi)=e^{2\lambda \phi} \ . 
\end{equation}
This choice corresponds to assuming that a single exponential term dominates in the regime of interest
(in the same spirit of the scaling solutions proposed, for instance,  in \cite{Gubser:2009qt}).
It is important to underline that the sign of $V$ in \eqref{pheno} is chosen to comply with the criterion 
of \cite{Gubser:2000nd} when the exponential blows up.

The metric ansatz for the solutions is taken to be%
\footnote{We fix the normalization of the spatial subspace volume to one.}
\begin{equation}\label{ans_met}
 ds^2=-a(r)^2dt^2+\frac{dr^2}{a(r)^2}+b(r)^2(dx_1^2+dx_2^2)\ ,
\end{equation}
with
\begin{equation}\label{ans_met2}
 a(r)=C_a r^{\gamma} \ , \qquad b(r)=r^{\beta} \ ,
\end{equation}
where $C_a$ is a numerical constant.
For the bulk gauge field we consider
\begin{equation}\label{ans_gau}
 F=\frac{Q}{f(\phi)\,b(r)^2}\, dt \wedge dr + B \, dx_1 \wedge dx_2\ ,
\end{equation}
corresponding in general to a dyonic bulk solution charged both electrically and magnetically.
The ansatz \eqref{ans_gau} solves the Maxwell equation coming from \eqref{model} automatically.
Eventually for the scalar we consider the following ansatz
\begin{equation}\label{ans_sca}
 \psi_I=k\, x_I \ , \qquad \phi=\kappa \log r\ .
\end{equation}
namely a logarithmic profile for $\phi$ and a linearly sourced (radially constant) 
scalar for each spatial direction whose linear spatial dependence is the source 
for momentum dissipation \cite{Andrade:2013gsa}.
In the dual perspective we have an electric charge density associated to a non-trivial $Q$ and
an ``off-plane'' magnetic field associated to $B$ (implemented as in, e.g., \cite{Hartnoll:2007ai}). 

The variable $r$ increases towards the boundary of the bulk. In the case of zero-temperature 
- corresponding to the absence of a horizon - there is no emblackening factor; therefore a naked 
singularity where $\phi$ diverges logarithmically appears in the deep IR. Its large value in the 
IR region constitute the rough motivation on which the hypothesis \eqref{pheno} relies; namely 
a single exponential term setting the behavior of the phenomenological functions. More comments are in order 
and we postpone them to the Discussion section. For now, let us just underline that 
\eqref{pheno} represents for us an IR hypothesis to build a low-energy effective holographic theory with scaling properties.
In fact, the solutions \eqref{ans_met}, \eqref{ans_gau} and \eqref{ans_sca} are not UV complete and would need a matching 
to an asymptotic UV AdS solution. For this to happen, also the form of the full potential must allow 
for a minimum at some finite value for $\phi$ to be reached asymptotically at large $r$.
 
The metric ansatz \eqref{ans_met} presents two parametric exponents $\gamma$ and $\beta$; as shown in Appendix \ref{hyperz},
they encode $\theta$ and $z$ as follows 
\begin{equation}
 \theta = \frac{2\,(1-\gamma)}{1-\beta-\gamma}\ , \qquad
 z = \frac{1 - 2\, \gamma}{1 - \beta - \gamma}\ .
 \label{zthetamap}
\end{equation}
We recall that $\theta$ and $z$ are associated respectively to the hyperscaling violation and the different 
scaling among the time and space directions of the field theory.
With \eqref{ans_met} we are therefore dealing with a hyperscaling violating Lifshitz geometry.

Given the phenomenological functions \eqref{pheno}, the ansatz \eqref{ans_sca}
and the consequent structure of the equations of motions, a concomitant flip of $\alpha$, $\delta$, $\lambda$ and $\kappa$
leaves everything unchanged (a part from an unimportant sign flip in the dilaton profile).
One can then restrict to 
\begin{equation}\label{del_pos}
 \delta > 0\ , 
\end{equation}
without losing in generality;
we take this choice in line with the analysis of \cite{Kundu:2012jn}.

\subsection{Equations of motion and effective potential}

The equations of motion for the model \eqref{model} and the metric ansatz \eqref{ans_met} read
\begin{align}
  \left(a^2\,b^2 \right)^{''}+2\,b^2\,V(\phi)+\frac{Y(\phi)}{2} \sum_I^2 (\partial_I \,\psi_I)^2 =0\ , \nonumber\\
  b^{''}+b\,\phi^{' \;2}=0\ , \nonumber\\
  \left(a^2\,b^2\phi'\right)'-\frac{1}{2}\partial_{\phi}V_{\text{eff}}(\phi)=0\ , \nonumber\\
  -\frac{1}{2}\,a^{2 \;'}\,b^{2 \;'}-a^2\,b^{' \;2}+a^2\,b^2\phi^{' \;2}-V_{\text{eff}}(\phi)=0\ . \label{EOM}
\end{align}
The Maxwell equation is trivially satisfied by the ansatz \eqref{ans_gau}.
The effective potential $V_{\text{eff}}(\phi)$ is given by
\begin{equation}\label{eff}
 V_{\text{eff}}(\phi)=\frac{Q^2}{b^2 f(\phi)}+\frac{B^2 \,f(\phi)}{b^2}+\frac{1}{2}\,b^2\,V(\phi)+\frac{Y(\phi)}{4}\sum_I^2 (\partial_I \,\psi_I)^2\ .
\end{equation}
For more details about the complete set of equations and their expressions on the ansatz (\ref{pheno}, \ref{ans_met2}, \ref{ans_sca}), we refer the reader to Appendix \ref{eom_ansa}.

\subsubsection{Scaling transformations}
\label{sca_sym}

The system admits the following scaling transformations (the same as in \cite{Kundu:2012jn})
\begin{eqnarray}
r & = & \lambda\ \tilde r \ , \label{rscale}\\
t & = & \lambda^{1-2 \gamma}\ \tilde t \ , \label{tscale}\\ 
x_I &=& \lambda^{1-\gamma-\beta}\ \tilde x_I \ ,\label{xyscale}
\end{eqnarray}
upon which the metric acquires a conformal factor
\begin{equation}
ds^{2} = \lambda^{2-2\,\gamma} \left\{ -C_a^2\, \tilde r^{2 \,\gamma}\, d \tilde
t^{2}+\frac{d\tilde r^{2}}{C_a^2\, \tilde r^{2 \,\gamma}}
+ \tilde r^{2 \,\beta} \left(d\tilde x^{2}_1 +d \tilde x^{2}_2\right) \right\}\ ;
\end{equation}
the dilaton is instead shifted
\begin{equation}
 \phi(r) = \kappa \log(\tilde r)+ \kappa \log(\lambda) = \phi(\tilde r) + \kappa \log(\lambda)\ .
\end{equation}

This scaling behavior is a property of the IR solutions. Namely, the scaling maps IR solutions into IR
solutions with rescaled parameters. To specify the scaling of the parameters 
one considers the various terms in the equations of motion (referring to the explicit form of $V_\text{eff}(\phi)$) 
and requires that they rescale in the same way. Later we will rely on this to fix the leading corrections 
to the thermodynamics due to irrelevant perturbations.

\subsection{Warm-up}
\label{wu}

A finite temperature can be realized modifying the $g_{tt}$ 
component of the metric introducing an emblackening factor,
\begin{equation}
\label{enb}
a^2(r)=C_a^2\,r^{2 \gamma} \left[1-\left(\frac{r_h}{r}\right)^{2 \beta +2 \,\gamma -1}\right] \ ,
\end{equation}
where $r_h$ is the horizon radius of the black hole. 
Following \cite{Kundu:2012jn},
the specific form of the emblackening factor \eqref{enb} is obtained by asking that the infrared equations 
of motion of the zero-temperature case are not modified by the introduction of a black hole horizon.
The black hole temperature is
\begin{equation}\label{te}
T = \left.\frac{(a^2)'}{4\pi}\right|_{r=r_h} = \frac{C_a^2}{4 \,\pi}\ (2\, \beta +2 \,\gamma -1)\ r_h^{2 \,\gamma -1}\ .
\end{equation}
Having assumed that the introduction of a small temperature does not affect the extremal geometry, 
the temperature must be small in comparison to the other scales of the system. 
We will be more precise on this point in discussing the thermodynamics of the solutions.

Finally, the entropy of the black hole is given as usual by the area law, namely
\begin{equation}\label{en}
S \propto T^{\frac{2\,\beta}{2 \,\gamma -1}} \propto T^{\frac{d_\theta}{z}} \ ,
\end{equation}
where $d_{\theta}$ is the effective dimension of the dual theory given by
\begin{equation}
 d_\theta = 2 - \theta \ .
\end{equation}

\subsection{DC transport}

The analytic computation of the DC transport coefficients relies on the ``membrane paradigm''
(in its generalized version to momentum-dissipating systems),
namely the fact that the zero-frequency currents are radially conserved \cite{Iqbal:2008by,Donos:2014cya,Amoretti:2014mma}.
Notice that this plays a particularly important r\^ole in a circumstance - like the present one - 
where one does not dispose of a complete knowledge of the backgrounds.

We follow closely the computations of \cite{Blake:2015ina} to which we refer%
\footnote{Similar analyses are present also elsewhere in the literature \cite{Amoretti:2015gna,Kim:2015wba,Zhou:2015dha}}.
The full set of DC transport coefficients reads
\begin{align}\label{sigma}
 &\sigma_{xx}=\frac{k^2\,b^2\, Y\left(B^2\, f^2+k^2\, b^2\, Y\, f+q ^2\right)}{\Delta}\Big|_h & \sigma_{xy}=\frac{B \,q\, \left(B^2 \,f^2+2\, k^2\, b^2\, Y f+q ^2\right)}{\Delta}\Big|_h\\ \label{alpha}
 &\alpha_{xx}\,=\,\frac{k^2 \,q\,  S \,b^2\, Y}{\Delta}\Big|_h \ \ & \ \ \alpha_{xy}\,=\,\frac{B\, S\, \left(B^2 \,f^2+k^2 \,b^2\,Y\, f+q^2\right)}{\Delta}\Big|_h\\ \label{kappa}
 &\bar{\kappa}_{xx}\,=\,\frac{S^2\, T\, \left(B^2 \,f+k^2 \,b^2\, Y\right)}{\Delta}\Big|_h \ \ & \ \ \bar{\kappa}_{xy}\,=\,\frac{B \,q\,  S^2\, T}{\Delta}\Big|_h
\end{align}
where the subscript $h$ means that the quantities are evaluated at the horizon; we have defined 
\begin{equation}
 \Delta = (B^2\,f+k^2\, b^2\,Y)^2+B^2\,q^2\ ,
\end{equation}
in analogy with \cite{Donos:2015bxe} and we have also indicated the charge density with $q$ and the entropy density with $S$;
these take the form
\begin{equation}
 q = - f\, b^2\, A_t'\ ,\qquad S = 4 \,\pi\, b^2 |_h\ ,
\end{equation}
accordingly to the metric \eqref{ans_met}. To show that $q$ is the charge density 
one first notices that $q$ is radially conserved because of the Maxwell equation and 
- when evaluated at a would-be $AAdS$ boundary - the two quantities would coincide.

From \eqref{sigma} one can derive the resistivity and the Hall angle obtaining
\begin{align}
 \rho_{xx} &= \left. k^2\, b^2\, Y \, \frac{4\,k^2\, b^2 \,f\, Y+16\, B^2\, f^2+q^2}{16 \,f^2 \left(k^4\, b^4\, Y^2+B^2 \,q^2\right)+8\, k^2\, q^2\, b^2\, f\, Y+q^4}\right|_h\ , \label{resistivity}\\
 \tan \theta_H &= \left.\frac{B \,q}{k^2\, b^2\, Y} \, \frac{8 \,k^2\, b^2\, f\, Y + 16 \,B^2\, f^2+q^2}
 {4\,k^2\,   b^2\, f\, Y+16\, B^2\, f^2+q^2}\right|_h\ .
 \label{cupratestransport}
\end{align}
Inverting \eqref{te} we have
\begin{equation}
 r_h = \left(\zeta \,T\right)^{\frac{1}{2\gamma - 1}}\ , 
\end{equation}
where we have defined
\begin{equation}
 \zeta = \frac{4\,\pi}{C_a^2} \, \frac{1}{2\, \beta + 2\, \gamma - 1}\ ,
\end{equation}
to avoid clutter.

\section{Magnetic field as an irrelevant perturbation}

In this section we focus on solutions where the magnetic field is an irrelevant perturbation 
that does not alter the nature of the IR fixed point described by the unperturbed gravitational solution. 
In particular, we are concerned with two types of solution: charged solutions where the electric 
charge density is marginal, and neutral solutions where both the charge density and the magnetic field
are irrelevant perturbations%
\footnote{These solutions have been already considered in the literature, see \cite{Gouteraux:2014hca,Gouteraux:2011ce,Donos:2014uba,Charmousis:2010zz}.}.

\subsection{Charged solutions}
 \label{csol}
In order to search for purely electric solutions, we set $B$ to zero in \eqref{ans_gau}. 
We require $k$, $\alpha$ and $V_0$ to be non-vanishing while $\delta$ to be positive as chosen 
in \eqref{del_pos}. Referring to the effective potential \eqref{eff} on the ansatz, we impose that 
the exponents of the three remaining terms (the magnetic term vanishes as $B=0$) are equal.
This amounts to requiring that the operators associated to $Q$ and to the scalars $\psi_I$
are marginal. The remaining parameters are fixed by the equations of motion \eqref{EOM} to be
\begin{equation}\label{soluzzo}
\begin{split}
 & \qquad  \qquad  \qquad  \qquad \qquad  \kappa=-\frac{2\, (\alpha +\delta )}{(\alpha +\delta )^2+4}\ , \qquad
  \lambda =\frac{\delta -\alpha }{2} \ , \\
 &\qquad \qquad \gamma = 1 + \delta\, \kappa = 1-\frac{2 \,\delta  \,(\alpha +\delta )}{(\alpha +\delta )^2+4} \ , \qquad
 \beta = 1- \gamma +\kappa \,\lambda = \frac{(\alpha +\delta )^2}{(\alpha +\delta )^2+4}\ .
 \end{split}
\end{equation}
along with
\begin{align}\label{Qcha}
Q^2 &=\frac{k^2 \left(-\alpha ^2+\delta ^2-4\right)-2 \delta  \left| V_0 \right| (\alpha +\delta )+4 |V_0|}{4\,
   [\alpha \, (\alpha +\delta )+2]} \ , \\ \label{Ccha}
C_a^2 &=-\frac{\left[(\alpha +\delta )^2+4\right]^2 \left(k^2-2 \left| V_0 \right| \right)}{4 \,[\alpha \, (\alpha +\delta
   )+2] \,[(3 \,\alpha -\delta ) \,(\alpha +\delta )+4]}\ .
\end{align}
The limit $k\rightarrow 0$ coincides with the solution found in \cite{Kundu:2012jn}.

Since $\beta$ in \eqref{soluzzo} is manifestly positive, requiring a positive specific heat from \eqref{en} 
corresponds to 
\begin{equation}
 \gamma > \frac{1}{2}\ .
\end{equation}
We also have to demand that the squared parameters given by \eqref{Qcha} and \eqref{Ccha}
are positive. We factorize $|V_0|$ and define $\xi = k^2 / |V_0|$; with simple manipulations we get
\begin{align}
 2 + \alpha\,(\alpha+\delta) > 0\ , \nonumber\\
 2 - \delta\,(\alpha+\delta) > 0\ , \nonumber\\
 4 + (3\,\alpha-\delta)\,(\alpha+\delta) > 0\ , \nonumber\\
 4 + (\alpha-3\,\delta)\,(\alpha+\delta)> 0\ ,
 \label{chargeregime}
\end{align}
and%
% \footnote{This last inequality could possibly provide diffusion bounds on the lines of \cite{Amoretti:2014ola}; 
% further study in this direction constitutes an interesting future perspective.}
\begin{equation}
 0 \leq \xi < 2\; \frac{2-\delta\,(\delta+\alpha)}{4+\alpha^2-\delta^2}< 1\ .
\end{equation}
Considering the vanishing $k$ limit we recover again the results described in \cite{Kundu:2012jn} of which the present solution represents a finite $k$ generalization.
The validity domain \eqref{chargeregime} satisfies the null energy conditions.
Relying on the map \eqref{zthetamap} and relations \eqref{soluzzo}, 
one can express \eqref{chargeregime} in terms of the physical quantities $z$ and $\theta$,
\begin{align}
 z<0 \ \wedge\ \theta> 2+z & \qquad  \text{or}\qquad\nonumber\\
 1<z\leq 3 \ \wedge\ \theta <2z-2 & \qquad \text{or}\qquad\nonumber\\ \label{domiz}
 z>3 \ \wedge\ \theta<1+z & 
\end{align}
and 
\begin{equation}
0\,\leq\,\xi\,<\,\frac{2 \,(z-1)}{2\, z-\theta }\ .
\label{xiregime}
\end{equation}
It is important to stress that in these new variables one has to additionally require $\kappa^2>0$,
after which the null energy conditions (see \cite{Gath:2012pg} for a derivation) 
\begin{align}\label{nec}
 (z-1)(z+2-\theta)\geq 0\ , \\
 (\theta-2)(\theta+2-2\,z)\geq 0\ , \label{deusiema}
\end{align}
are automatically satisfied.

\begin{figure}
\centering
\includegraphics[width=7.5cm]{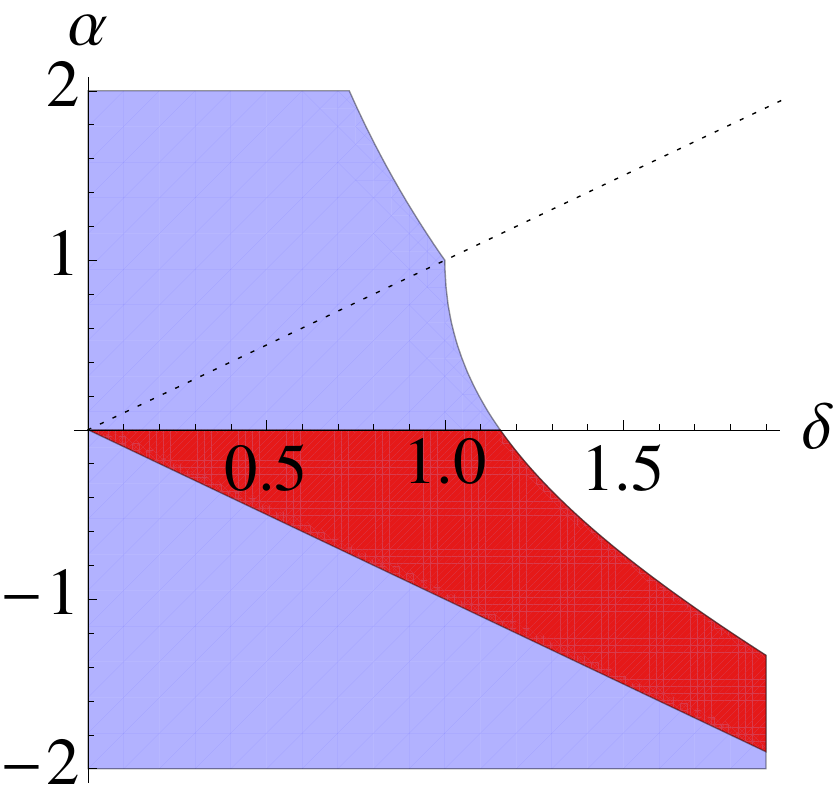}%
\qquad
\includegraphics[width=7cm]{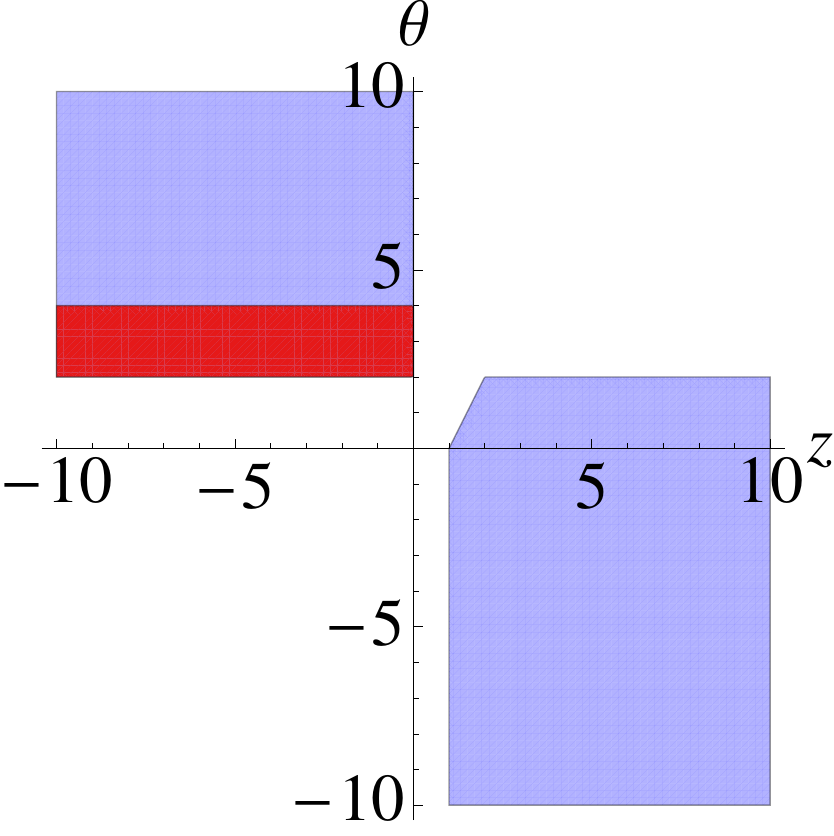}
\caption{The colored areas correspond to the domain of validity of the charged solution depicted 
in the $(\delta,\alpha)$-plane (\emph{left}) and in the $(z,\theta)$-plane (\emph{right}).
The red areas represent the restriction to the region where $B$ corresponds to an irrelevant operator.
The dotted line corresponds to $\alpha=\delta$.}\label{domino}
\end{figure}

\subsubsection{Perturbing with a magnetic field}

We perturb the charged solutions \eqref{soluzzo} by means of a small magnetic field $B$ and 
focus on the case where it is dual to an irrelevant operator. First, we compare 
the various terms in the effective potential $V_{\text{eff}}(\phi)$ in \eqref{eff} and ask for the 
term in $B$ to be small compared to all the other terms, namely
\begin{equation}
 \frac{B^2}{Q^2} \ll e^{-4\alpha\phi}\ , \qquad
 \frac{B^2}{k^2}   \ll \frac{b^2}{2} e^{2(\lambda-\alpha)\phi}\ , \qquad
 \frac{B^2}{|V_0|} \ll \frac{b^4}{2} e^{-2 \phi(\alpha-\delta)}\ .
\end{equation}
Considering the ansatz \eqref{ans_met}, we obtain
\begin{equation}\label{condizie}
 \frac{B^2}{Q^2} \ll r^{-4 \alpha \kappa} \ , \qquad 
 \frac{B^2}{k^2}   \ll \frac{r^{-4 \alpha \kappa}}{2} \ , \qquad 
 \frac{B^2}{|V_0|} \ll \frac{r^{-4 \alpha \kappa}}{2}\ ;
\end{equation}
these encode the fact that $B$ is a perturbation. The purely charged solution \eqref{soluzzo} can be considered as a reliable solution of the Einstein equations
in a range of the radial coordinates where the conditions \eqref{condizie} are satisfied.
The appearance of the same exponent in all three conditions \eqref{condizie} is a direct consequence of the marginality 
of $Q$, $k$ and $V$.

We focus on cases where the magnetic field is an irrelevant perturbation and 
\eqref{condizie} are satisfied all the way down to the IR, namely we consider 
\begin{equation}
 \alpha \,(\alpha+\delta) < 0\ ,
 \label{bullo}
\end{equation}
which corresponds to $\alpha\,\kappa>0$ upon using \eqref{soluzzo}.
Translating this condition in terms of $\theta$ and $z$, one obtains
\begin{align}
z\leq 2\ \ \land\ \ 2<\theta <4   & \qquad  \text{or}\qquad\\
2<z<3\ \ \land\ \ 2 z-2<\theta <4 & \ .
\end{align}
This, once intersected with the regime of validity of the charged solution \eqref{domiz}, yields
\begin{equation}
z<0\ \ \land\ \ 2<\theta<4\ ;
\label{chargedregime3}
\end{equation}
to be considered together with \eqref{xiregime}.
The validity domain of the charged solution where $B$ corresponds to a dual irrelevant operator is shown in red in Figure \ref{domino}.

\subsubsection{Thermodynamics}

To understand the interplay of the different scales in the charged solution \eqref{soluzzo},
it is important to discuss the thermodynamic corrections due to the introduction of the 
$B$ perturbation.
To this end we rely on the scaling symmetry \eqref{rscale}, \eqref{tscale}, \eqref{xyscale} of the IR solution. 
Since $\mu$ and $k$ are UV quantities, they do not rescale.
Notice that for $k$ the statement is 
particularly non-trivial because $k$ appears also as a parameter in the IR. We come back to this shortly.

From \eqref{en} and dimensional analysis we have that the entropy density $S$ has to scale as follows:
\begin{equation}
 S \propto \bar \mu^2\, \left(\frac{T}{\bar \mu}\right)^{\frac{2\beta}{2\gamma-1}} \ .
\end{equation}
The UV scale $\bar \mu$ is defined by
\begin{equation}\label{UVS}
 \bar \mu = \mu\  h\left(\frac{k}{\mu}\right)\ ,
\end{equation}
where $\mu$ is the chemical potential, $k$ the momentum dissipation scale and $h$ is a dimensionless 
function of the dimensionless ratio $k/\mu$ which is analytical in $k$ and satisfies 
\begin{equation}
 h(k/\mu) \rightarrow 1 \ \qquad \text{for} \qquad k\rightarrow 0\ ;
\end{equation}
this last condition is imposed asking that we reduce to the analysis of \cite{Kundu:2012jn} when we switch off the momentum dissipation.

Since $B$ enters always quadratically in the equations of motion, 
the first correction to the entropy has to be quadratic in $B$. Such leading correction term must be dimensionless and scaling invariant.
We have first to determine the scaling of $B$, which is derived asking that all the terms in $V_{\text{eff}}(\phi)$ scale as the other 
terms of the equations of motion where $V_{\text{eff}}(\phi)$ appears. From this we have that upon rescaling
\begin{equation}
 B \rightarrow B\,  r^{2\beta +\kappa(\delta-\alpha)} = B\, r^{-2\alpha \kappa}\ ,
\end{equation}
where we have used relations \eqref{soluzzo} among the exponents. 
Eventually, the correction to the entropy density due to the introduction of $B$ are given by
\begin{equation}
 S \propto  \bar \mu^2\, \left(\frac{T}{\bar \mu}\right)^{\frac{2\beta}{2\gamma-1}} 
 \left[1 + s_1 \left(\frac{B}{\bar \mu^2}\right)^2 \left(\frac{T}{\bar \mu}\right)^{\frac{4\alpha \kappa}{2\gamma-1}}  \right]\ ,
\end{equation}
where $s_1$ is a constant parameter whose determination would require 
the analysis of the backreaction of $B$ on the geometry.

Finally, the  condition for the entropy correction to be small is given by:
\begin{equation}
\label{finalcond}
\left(\frac{T}{\bar \mu} \right)^{\frac{4 \alpha \kappa}{2 \gamma-1}}\frac{B^2}{\bar \mu^4} \ll 1 \ .
\end{equation} 
It is important to note that, due to the irrelevance of the magnetic perturbation, 
the exponent ${\frac{4 \alpha \kappa}{2 \gamma-1}}$ is always positive. 
As long as the condition \eqref{finalcond} is satisfied, one can 
consider either a regime where $T$ is small compared to the magnetic field $B$
or the other way around.

\subsubsection{DC transport on the charged solutions}
\label{tra_cha}

In this section we discuss the temperature scalings of the longitudinal electric resistivity and Hall angle
on the charged solutions \eqref{soluzzo}. 
Inserting the phenomenological functions \eqref{pheno} into expressions \eqref{cupratestransport} and working up to linear order in $B$, we get
\begin{equation}\label{rocchio}
 \rho_{xx} =  \frac{k^2\, (\zeta \,T)^{2 \frac{\beta + \kappa \lambda}{2 \,\gamma - 1}}}{q^2 + 4 \,k^2\, (\zeta\, T)^{2\frac{\beta + \kappa\, (\alpha + \lambda)}{2\, \gamma - 1}}} + {\cal O}(B^2)\ ,
\end{equation}
and
\begin{equation}\label{Hocchio}
 \tan \theta_H = \frac{ B\, q}{k^2}\ (\zeta \,T)^{-2\frac{\beta + \kappa \,\lambda}{2\, \gamma - 1}}\ \frac{q^2 + 8\, k^2 (\zeta \,T)^{2\,\frac{\beta+\kappa\,(\alpha+\lambda)}{2 \,\gamma - 1}}}
 {q^2 + 4\,k^2\, (\zeta\, T)^{2\frac{\beta+\kappa(\alpha+\lambda)}{2 \,\gamma - 1}}} + {\cal O}(B^2)\ .
\end{equation}

Using the relations coming from the charged background solution \eqref{soluzzo} to rewrite the exponents in \eqref{rocchio} and \eqref{Hocchio}, 
\begin{equation}
 2\, \frac{\beta + \kappa(\alpha+\lambda)}{2\,\gamma - 1} = 0\ , \qquad 
 2\, \frac{\beta + \kappa \,\lambda}{2\,\gamma - 1} = \frac{4 \alpha (\alpha + \delta)}{4 + (\alpha - 3 \delta)(\alpha + \delta)}\ ,
 \label{constcharged}
\end{equation}
we obtain 
\begin{equation}\label{sca_rho_fl}
 \rho_{xx} = \frac{k^2 }{q^2 + 4\,k^2 } (\zeta\, T)^{\frac{4\, \alpha (\alpha + \delta)}{4 + (\alpha - 3 \delta)(\alpha + \delta)}} + {\cal O}(B^2)\ ,
\end{equation}
and
\begin{equation}\label{sca_hal_fl}
 \tan \theta_H = \frac{B\, q}{k^2}\ \frac{q^2 + 8 \,k^2 }{q^2 + 4\,k^2 }\ (\zeta\, T)^{-\frac{4 \,\alpha (\alpha + \delta)}{4 + (\alpha - 3\, \delta)(\alpha + \delta)}}  + {\cal O}(B^2)\ .
\end{equation} 

It is crucial to note that the scalings of the transport coefficients arising from the class of charged solutions perturbed with an 
irrelevant magnetic field are of the \textit{Fermi liquid} type, namely 
\begin{equation}
\rho_{xx}\,\propto\,T^n\,,\,\,\,\,\,\tan \theta_H\,\propto\,T^{-n}\,.
\end{equation} 
Within this class of solutions there is no room for a \textit{non-Fermi liquid} 
behavior and specifically for a holographic accommodation of the cuprate scalings, 
\begin{equation}
\rho_{xx}\,\propto\,T\,,\,\,\,\,\,\tan \theta_H\,\propto\,T^{-2}\,.
\end{equation} 

It is interesting to discuss whether the solutions in the charged class provide
a metallic or an insulating behavior. To this purpose, we refer to the following definitions%
\begin{align}
&\textit{metal}\,\,:\;\; \frac{d\rho_{xx}}{dT}\,>\,0\,,\qquad\qquad\textit{insulator}\,\,:\;\; \frac{d\rho_{xx}}{dT}\,<\,0\,.
\label{metaldef}
\end{align}
From \eqref{sca_rho_fl} we have that the resistivity in this class scales like $\rho_{xx}\propto T^{m_1}$ with:
\begin{equation*}
m_1\,=\,\frac{4\, \alpha (\alpha + \delta)}{4 + (\alpha - 3 \delta)(\alpha + \delta)}\,<\,0\,.
\end{equation*}
Therefore this class of solutions corresponds to insulating systems%
\footnote{This statement holds only if one insists on an irrelevant magnetic field. 
Allowing for the magnetic field to be marginal or relevant might change the picture.}
and the hunt for the \textit{strange metal} phenomenology has to consider other setups.

\subsection{Neutral solutions}
\label{nsol}

The analysis of the transport properties of the charged solutions described in Subsection \ref{tra_cha}
shows that the temperature scaling of $\rho$ \eqref{sca_rho_fl} and $\cot \theta_H$ \eqref{sca_hal_fl} is the same. 
This is due to the fact that the electric conductivity is dominated by the momentum-dissipation 
physics
which determines the scale of $\cot \theta_H$ as well.

In order to have different temperature scalings for $\rho$ and $\cot \theta_H$ we consider a geometry 
in which both the charge density and the magnetic field are perturbative (and irrelevant). The 
momentum-dissipation term in the longitudinal conductivity is controlled by the square of the parameter 
$Q$ which is now perturbative. 

It is important to stress that, even though the solution at hand allows for the possibility 
of considering both $B$ and $Q$ of the same order of magnitude, we restrict only to cases 
where $Q \gg B$. This choice is in line with generic experimental circumstances.

We solve Einstein equations \eqref{eom_ansa} with both $B$ and $Q$ set to zero
obtaining 
\begin{equation}\label{neutero}
 \begin{split}
 \kappa = & \frac{\lambda - \delta}{(\delta-\lambda)^2+1}\ , \qquad 
 \gamma = 1 + \kappa \delta = 1 + \frac{\delta(\lambda - \delta)}{(\delta-\lambda)^2+1}\ , \\
 &\qquad \beta = 1 - \gamma + \kappa\lambda = \frac{(\delta-\lambda)^2}{(\delta-\lambda)^2+1}\ ,
 \end{split}
\end{equation}
and 
\begin{align}
 C_a^2 &= \frac{\left|V_0\right|\left((\delta -\lambda )^2+1\right)^2}
 {2 [1-\lambda(\delta-\lambda)] \left(\delta ^2-4 \delta  \lambda +3 \lambda ^2+1\right)}\ ,\\ \label{ksol}
k^2 & = |V_0| \frac{\delta  (\lambda -\delta )+1}{\lambda^2-\delta\lambda+1}\ .
\end{align}
The domain of validity of these neutral solutions is obtained by demanding the positivity 
of the squared quantities $C_a^2$ and $k^2$ along with the positivity of 
the specific heat \eqref{en}; this leads to
\begin{align}
 1-\lambda(\delta-\lambda)>0\ ,\nonumber\\
 1+\delta(\lambda-\delta)>0\ ,\nonumber\\
 1+(\lambda+\delta)(\lambda-\delta)>0\ ,\nonumber\\
 1+(\delta-3\lambda)(\delta-\lambda)>0\ .
 \label{neutralregime}
\end{align}
We still refer to the choice $\delta>0$ commented around Equation \eqref{del_pos}.

The domain of validity of the neutral solutions \eqref{neutralregime} can be expressed in terms of $\theta$ and $z$
and, as in the charged case, one has to consider explicitly the null energy conditions \eqref{nec}; 
eventually one obtains for the neutral solutions the same validity domain on the $(z,\theta)$-plane
as for the charged solutions \eqref{soluzzo}, namely
the colored region (both blue and red areas) of the right plot in Figure \ref{domino}.

\subsubsection{Perturbing with a magnetic field and a charge density}

The irrelevance of the magnetic field and and the charge density means that the 
corresponding terms in $V_{\text{eff}}(\phi)$ \eqref{eff} become smaller and smaller towards the IR faster 
than the pure dilatonic term.
In addition, $Q$ and $B$ can be considered to introduce small perturbations of the neutral solutions \eqref{neutero}
in the radial interval where the following conditions are satisfied:
\begin{equation}\label{condQe}
 \frac{B^2}{k^2}   \ll \frac{1}{2} r^{2\beta+2(\lambda-\alpha)\kappa}\ , \qquad
 \frac{B^2}{|V_0|} \ll \frac{1}{2} r^{4 \beta-2 \kappa(\alpha-\delta)}\ .
\end{equation}
and 
\begin{equation}\label{condQm}
 \frac{Q^2}{k^2}   \ll \frac{1}{2} r^{2 \beta+2(\lambda+\alpha)\kappa}\ , \qquad
 \frac{Q^2}{|V_0|} \ll \frac{1}{2} e^{4 \beta+2 \kappa(\alpha+\delta)}\ .
\end{equation}
Again, the case where both the electric and the magnetic perturbations 
are irrelevant corresponds to having the conditions \eqref{condQe} and 
\eqref{condQm} valid all the way down to the IR%
\footnote{In \eqref{condQm} and \eqref{condQe} one obtains the same $r$ exponent
when comparing with both $k^2$ and $|V_0|$ (one can appreciate this explicitly upon using 
relations \eqref{neutero}). This fact descends directly form the marginality 
of both the momentum-dissipating scalars and the dilaton.}.
Explicitly, the irrelevance requirements correspond to
\begin{align}
 (\lambda-\delta)(2\lambda-\delta+\alpha)<0\ ,\nonumber \\
 (\lambda-\delta)(2\lambda-\delta-\alpha)<0\ .
 \label{neutralboh}
\end{align}

\subsubsection{Thermodynamics}

The analysis of the $Q$ and $B$ corrections to the thermodynamics (specifically, to the entropy density)
instructs us about the relations among the different physical scales of the system and about the regime of validity
of the perturbative approximations.

The analysis is analogous to that performed already for the charged solutions. 
At zero order in $Q$ and $B$, the temperature scaling of the entropy density is obtained combining 
the ansatz for the unperturbed solution \eqref{ans_met} with the relation connecting the temperature and the horizon radius \eqref{te}; 
one obtains
\begin{equation}
\label{entr1}
 S \propto \bar\mu^2 \left(\frac{T}{\bar\mu}\right)^{\frac{2\beta}{2\gamma - 1}}\ .
\end{equation}
The appropriate insertions of $\bar\mu$ (see \eqref{UVS} and comments around it) follow from dimensional considerations.

Relying upon the scaling symmetry \eqref{sca_sym} and enforcing that all the terms in $V_{\text{eff}}(\phi)$ scale as the other 
terms of the equations of motion where $V_{\text{eff}}(\phi)$ appears, we obtain that both $Q$ and $B$ have to scale as follows:
\begin{align}\label{Qmsca}
 B \rightarrow B\: r^{2\beta+ \kappa(\delta-\alpha)} &= B\: r^{-\kappa(\alpha+\delta-2\lambda)} \ ,\\ \label{Qesca}
 Q \rightarrow Q\: r^{2\beta+ \kappa(\delta+\alpha)} &= Q\: r^{\kappa(\alpha-\delta+2\lambda)} \ .
\end{align}
The corrections to the entropy have to be quadratic in $Q$ and $B$ (as these quantities enter quadratically in the bulk equations of motion); 
relying on dimensional analysis and the scalings \eqref{Qmsca} and \eqref{Qesca}, 
we find the following form for the leading corrections to the entropy density \eqref{entr1}:
\begin{equation}\label{Scorr}
 S \propto  \bar\mu^2\, \left(\frac{T}{\bar\mu}\right)^{\frac{2\beta}{2\gamma-1}} 
 \left[1 + s_1 \left(\frac{B}{\bar\mu^2}\right)^2 \left(\frac{T}{\bar\mu}\right)^{2\kappa\frac{\alpha+\delta-2\lambda}{2\gamma-1}} 
 + s_2 \left(\frac{Q}{\bar\mu^2}\right)^2 \left(\frac{T}{\bar\mu}\right)^{-2\kappa\frac{\alpha-\delta+2\lambda}{2\gamma-1}} \right]\ ,
\end{equation}
where $s_1$ and $s_2$ are two numerical parameters which could be fixed 
considering the backreaction of $Q$ and $B$ on the background.
Finally, in order for both the correction terms in \eqref{Scorr} to be small perturbations, 
the various scales into play are constrained by the following relations:
\begin{equation}\label{condez}
 \left(\frac{B}{\bar\mu^2}\right)^2 \left(\frac{T}{\bar\mu}\right)^{2\kappa\frac{\alpha+\delta-2\lambda}{2\gamma-1}} \ll 1\ , \qquad 
 \left(\frac{Q}{\bar\mu^2}\right)^2 \left(\frac{T}{\bar\mu}\right)^{-2\kappa\frac{\alpha-\delta+2\lambda}{2\gamma-1}} \ll 1 \ .
\end{equation} 
Similarly to what obtained in \eqref{finalcond}, due to the irrelevance of the perturbations, 
the exponents of $T/\bar\mu$ appearing in both the previous relations are positive;
analogous comments made there hold here as well.

\subsubsection{DC transport on the neutral solutions}

We now analyze the transport properties of the neutral solution \eqref{neutero}. 
The electric resistivity and the Hall angle \eqref{cupratestransport} on this solution take the form
\begin{equation}
 \rho_{xx} = \frac{1}{4} (\zeta\, T)^{\frac{2 \,\alpha\,  (\lambda - \delta)}{\delta ^2-\lambda^2-1}}\,,\qquad
 \tan \theta_H = \frac{2 \,B\, q}{k^2} (\zeta\, T)^{\frac{2 (2\lambda - \delta) (\lambda- \delta)}{\delta ^2-\lambda ^2-1}}\ .
\end{equation} 
We recall that, since both the charge density 
and the magnetic field are perturbations, we are working up to linear order in $q$ and $B$.
In particular, let us emphasize that the electric DC conductivity is no longer dominated by the momentum-dissipating term; 
this feature leaves room for a possible \textit{non-Fermi liquid} behavior.
As already anticipated, in this scenario the scalings of the resistivity $\rho_{xx}$ and the 
Hall angle $\cot \theta_H$ are \textit{a priori} different. 
Nevertheless, demanding the two scalings to fit the cuprate behavior, namely
\begin{equation}
 \frac{2 \,\alpha\,  (\lambda - \delta)}{\delta ^2-\lambda^2-1} = 1\ , \qquad
 \frac{2 (2\lambda - \delta) (\lambda- \delta)}{\delta ^2-\lambda ^2-1} = -2\ ,
\end{equation}
one finds no solution that lie inside the domain of the neutral backgrounds (\ref{neutralregime}, \ref{neutralboh}).
In conclusion, despite the possibility of having a \textit{non-Fermi liquid} behavior in the class of neutral solutions, 
it is anyhow not possible to accommodate the cuprate phenomenology .

The scaling of the electric resistivity within this class goes like $\rho_{xx}\propto T^{m_2}$ where:
\begin{equation*}
m_2\,= \frac{2 \,\alpha\,  (\lambda - \delta)}{\delta ^2-\lambda^2-1}\ .
\end{equation*}
Being the denominator negative definite on the regime of definition of the solutions, the requirements of having a metal or an insulator translate into:
\begin{align*}
&\textit{metal}\,\,:\;\; \alpha\,(\delta-\lambda)\,<\,0\,,\qquad\qquad\textit{insulator}\,\,:\;\; \alpha\,(\delta-\lambda)\,>\,0\,.
\end{align*}
Both the conditions have non null overlap with the regime of validity \eqref{neutralregime} and the conditions \eqref{neutralboh} for the
irrelevance of the perturbations; within this class both metallic and insulating behaviors are then allowed.

\section{Discussion}
\label{discu}

\subsection{Other backgrounds}
\label{other}
The analysis so far has delved into two backgrounds where the momentum-dissipating 
scalars were always dual to marginal deformations of the quantum field theory.
Moreover, the magnetic field has always been considered as an irrelevant perturbation.
Some comments on other setups and their physical features is in order.

First, one can consider cases where the momentum-dissipating scalars are perturbations dual to irrelevant 
operators. The unperturbed background can be either charged or neutral.
The analysis proceeds similarly to the cases presented explicitly above however, being now the momentum-dissipating 
device perturbative, the leading order of the corrections to the thermodynamics is $k^2$ (to be seen 
as the counter-parts of the $B^2$ or $Q^2$ corrections discussed above). The computation of the transport 
coefficients yields a longitudinal resistivity which goes as $k^2$ at leading order. This is however 
the order at which the momentum-dissipating scalars would back-react on the background.
Before performing a deeper perturbative analysis, it is therefore unclear whether this result concerning the resistivity 
could be relied upon.

Another possibility is to consider a magnetically charged background (with marginal momentum-dissipating 
scalar); this being in spirit the electro-magnetic dual of the charged solution analyzed above.
The analysis of such a setup does not present technical obstructions, nevertheless is arguably at odds with 
typical experimental circumstances as it would feature a magnetic field dominating over all the other scales 
along the entire portion of renormalization flow described by the holographic background%
\footnote{More precisely, a marginal $B$ could be comparable with the momentum-dissipating sector but 
dominating by construction over a perturbative charge density and temperature.}.
These cases, which are physically relevant in general, appear not to be interesting in view of 
describing the cuprate phenomenology, they do not accommodate the right scalings in $T$ of 
the resistivity and Hall angle.

\subsection{UV}

All the solutions presented in this paper are based on the hypothesis that the field $\phi$ 
is marginally relevant and that its absolute value becomes large at the horizon.
To the aim of having an IR effective picture, we then assumed that the complete potential 
for $\phi$ is dominated by one single term \eqref{pheno}; therefore \eqref{model} 
is to be regarded as in IR effective theory, namely a deep bulk model.

The proper definition of the field/operator map requires in general a precise knowledge
of the UV behavior of the fields and a full-fledged holographic renormalization procedure.
Our solutions and even our bulk Lagrangian \eqref{model} are however not physically 
trustworthy in the UV%
\footnote{A systematic program of Lifshitz holographic 
renormalization is described in \cite{Chemissany:2014xsa,Papadimitriou:2014lia,Taylor:2015glc}.}.
To amend for this, we could in principle resort to a UV completion where 
the full potential for $\phi$ features minima at finite field values and admits $AAdS$ 
solutions. In other words, one could repeat a program along the lines of \cite{Kundu:2012jn,Harrison:2012vy} 
extended to comprehend the momentum-dissipating scalars.
Instead of investigating this possibility of completing or matching the IR 
solution (and theory) with some UV counterpart, we assume $AdS$ asymptotics
and argue about the insensitivity of our results to the UV detail%
\footnote{Relevant comments on similar issues are contained in the recent paper \cite{Gursoy:2015nza}.}. 

A fundamental property of the DC currents that has been considered here
consists in being radially constant in line with a generalized version 
of the ``membrane paradigm'' \cite{Iqbal:2008by,Donos:2014cya,Amoretti:2014mma}.
Such crucial feature should be respected in a would-be UV completion.
Suppose that we completed the theory in the UV by means of a bigger model 
admitting $AAdS$ solutions. Given the dimensionality of interest, it is known 
that both the electric current and the heat current are finite. 
They do not need to be renormalized and, importantly, there are no 
finite counter-terms leading to ambiguities in their definition \cite{Amoretti:2014zha}.
The same is true for the associated conductivities too.
So, on the basis of an assumed completion featuring a UV conformal fixed point \Daniele{(namely $AdS$ asymptotics)}
and the membrane paradigm, we can argue that the current operators that we define and their associated conductivities 
are the same that we would define in a complete theory with full knowledge of the solutions.

\subsection{IR}
\label{IR_dis}

The $T=0$ temperature solutions have a mild (namely, logarithmic) naked singularity
and one could question its acceptability. A definite answer to this would probably 
require a precise string completion of the bulk model. This is of course beyond the aim
of the present paper, so we resort to the standard ``bottom-up'' criteria 
described in \cite{Gubser:2000nd}: the boundedness of the potential from \emph{above},
and the possibility of obtaining the extremal solution as a zero-temperature limit of black-brane solutions
(what usually goes under the name of cloaking the singularity with a horizon)%
\footnote{The well-posedness of the Sturm-Liouville problem for the fluctuations is proposed as a further 
acceptability criterion for the background in \cite{Charmousis:2010zz,Kiritsis:2015oxa}; our solutions meet this criterion too.}.
As we have already noted, our solutions pass the former criterion because we take $V\sim-|V_0|e^{2\delta\phi}$ in \eqref{pheno}.
The second criterion
is passed as well; we obtained the nearly-extremal solutions finding a suitable one-parameter 
deformation of the $T=0$ solutions. Therefore, by construction, we have that the extremal solutions 
are obtained as the $T=0$ limit of the one-parameter family itself.

The acceptability (in string theory) of the singularity does not instruct us about how 
actually to resolve it. In other words, the approximation regarding the description by means 
of a classical and two-derivative gravity could breakdown close to the singularity.
Namely the IR problems could be of two kinds: large curvatures and quantum corrections.
These in turn could imply a temperature threshold below which the model is not reliable.
On this issues we borrow the comments on how to tame the singularity described in 
\cite{Iizuka:2011hg} to which we refer. In fact, the presence of additional scalars
does not change the qualitative bottom-up picture. The moral is that the large $N$ limit 
helps in controlling both issues even though conclusive answers could only be provided 
by precise string insight. Indeed the quantities involved in possible lower bounds for $T$, 
such as the four-dimensional Plank scale, are derived quantities. Namely, their definition 
relates to the original string model and involves non-trivially the higher dimensional fields
and dynamics thereof.

It is interesting to consider a direct approach to the problem of IR subtleties 
as in \cite{Harrison:2012vy} where additional terms in the gauge kinetic function 
(the $f(\phi)$ in \eqref{model}) are explicitly considered in the cases where 
the bulk gauge coupling becomes large. Such corrections drive the geometry away from the 
hyperscaling-violating-Lifshitz ansatz and furnish an IR completion to that. We do not 
enter into the detail of such IR completion which generically features $AdS$ 
factors and therefore residual entropies.  

All in all, the gravity description we adopt is likely to be safe from possible 
lower bounds for $T$ provided by either high curvatures and by a large bulk gauge coupling 
thanks to the large $N$ limit and the possibility to tune the UV value of $\phi$ \cite{Iizuka:2011hg,Harrison:2012vy}.
Any statement involving a strict $T\rightarrow0$ limit however should be taken with extreme care.
Notably, comments about the residual entropy fall in this class.

\section{Future perspectives}
\label{futu}

One important question that remains open is the individuation of a holographic model 
capable of reproducing the temperature behavior of the longitudinal resistivity and 
the Hall angle characterizing the cuprate physics. 
A possible continuation of the present analysis could be pursued 
considering also the cases where the charge density is a relevant operator in the IR;
this could potentially overcome the obstructions we pointed out and allow one to accommodate the cuprate scalings.
To the same purpose it could be inspiring to go back to the original proposal of \cite{androson}, where the insertion of 
an effective spin-spin interaction led to a description characterized
by two different scattering rates for longitudinal and transverse modes.
It is interesting to ask ourselves how to realize a similar proposal within 
the holographic context and, more specifically, in generalizations of the present setup. 
One could either try to actually embed some 
devices describing the spin dynamics%
\footnote{In this sense it it interesting to see whether the setups described in 
\cite{Bigazzi:2011ak,Musso:2013rva,Musso:2013rnr,Amoretti:2013oia} could provide a useful starting point.}
or (at least as a starting point) to just consider 
features which lead to an extra scale and a decoupling of the longitudinal and transverse physics.

Additionally, it could be interesting to think about Chern-Simons terms and their 
impact on IR scaling geometries. Another way to expand the range of results presented here would be to consider
two $U(1)$'s, one carrying the electric charge and the other the magnetic
field (along the lines of \cite{Gnecchi:2016auh}). This could relax some of the constraints encountered in the present analysis.

%Another interesting extension would be to generalize the present analysis 
%to take into account an anomalous dimension for the gauge field in line with \cite{Hartnoll:2015sea,Karch:2014mba}.
%One direct way of producing such an anomalous dimension for the gauge field is provided by the unparticles 
%models%
%\footnote{See \cite{Karch:2015pha,Karch:2015zqd} for the construction and the relations with the cuprate phenomenology.}.

Finally, it would be interesting to use the solutions here analyzed to address the investigation about 
bounds in momentum-dissipation holography, both for diffusion%
\footnote{We refer also to the line addressing disorder-driven metal-insulator transitions
\cite{Grozdanov:2015qia,Grozdanov:2015djs,Baggioli:2016oqk,Gouteraux:2016wxj,Fadafan:2016gmx}.}
\cite{Policastro:2001yc,Hartnoll:2014lpa,Kovtun:2014nsa,Amoretti:2014ola,Alberte:2016xja,Liu:2016njg,Burikham:2016roo}
and entropy production by a strain \cite{Hartnoll:2016tri}. In particular to systematically test the proposal of 
\cite{Hartnoll:2016tri} to relate the irrelevance of the momentum dissipation mechanism to a constant $\eta/S$ bounding 
value at zero temperature and eventually studying the temperature fall-off scaling of the $\eta/S$ ratio (see for example \cite{Ling:2016ien}).

\section{Acknowledgments}

Particular thanks go to Alessandro Braggio, Aldo Cotrone and David Tong for their feedback on the draft. 
We would also like to thank the JHEP referees for the work done in reviewing this paper.

DM would like to thank Daniel Arean, Riccardo Argurio, Matteo Bertolini, Piermarco Fonda, Andrea Mezzalira, Francesco Nitti, Victor Giraldo Rivera, 
Leopoldo Pando Zayas, Giuseppe Policastro, Antonello Scardicchio, Koenraad Schalm, Marika Taylor and Vipin Varma
for insightful and nice exchanges on the topics of this paper.

MB would like to thank Ren\'e Meyer, Elias Kiritsis and Oriol Pujol\'as for valuable discussions and comments about this work.

AA and DM would like to thank Carlo Baghino for inspiring encouragements.

AA is supported by the European Research Council under the European 
Union's Seventh Framework Programme (FP7/2007-2013), ERC grant agreement 
STG 279943, ``Strongly Coupled Systems''

\appendix

\section{Relation among $\gamma$, $\beta$ and $\theta$, $z$}
\label{hyperz}

Consider the generic form for a hyperscaling-violating Lifshitz geometry (see for instance \cite{Dong:2012se,Gouteraux:2014hca})
\begin{equation}\label{methyp}
 ds^2 = \bar r^{\, \theta} \left[- {\cal U}(\bar r) \frac{dt^2}{\bar r^{\, 2z}} + \frac{d\bar r^2}{{\cal U}(\bar r)\bar r^{\, 2}} + \frac{dx^2_1 + dx^2_2}{\bar r^{\, 2}}\right]\ .
\end{equation}
We consider the following radial change of coordinate
\begin{equation}
 r = \bar r^{\, \nu}\ , \qquad dr = \nu\; \bar r^{\, \nu - 1} d\bar r\ .
\end{equation}
We demand
\begin{equation}
 a[r(\bar r)]^2 = {\cal U}(\bar r)\bar r^{\, \theta - 2z}\ , \qquad
 \frac{dr^2}{a[r(\bar r)]^2} = \bar r^{\, \theta} \frac{d\bar r^2}{{\cal U}(\bar r) \bar r^{\, 2}}\ , \qquad
 b[r(\bar r)]^2 = \bar r^{\, \theta - 2}\ ,
\end{equation}
where the functions $a$ and $b$ are those appearing in the ansatz \eqref{ans_met}.
We obtain
\begin{equation}
 \bar r^{\, 2\gamma\nu} \sim \bar r^{\, \theta - 2z}\ , \qquad
 \bar r^{\, 2(\nu-1)} \sim \bar r^{\, \theta+2(\gamma\nu-1)}\ , \qquad
 \bar r^{\, 2\beta\nu} \sim= \bar r^{\, \theta - 2}\ ,
\end{equation}
from which we eventually get
\begin{equation}
 \nu = \frac{1}{1-\beta-\gamma}\ , \qquad 
 \theta = \frac{2(1-\gamma)}{1-\beta-\gamma}\ , \qquad
 z = \frac{1 - 2 \gamma}{1 - \beta - \gamma}\ .
\end{equation}
This is in agreement with similar computations performed in \cite{Kundu:2012jn}. Note however
that they referred to the hyperscaling metric as given in \cite{Huijse:2011ef}; this latter is mapped to \eqref{methyp}
by means of a further change of radial coordinate $r\rightarrow r^{(2-\theta)/2}$.

\section{Equations of motion on the ansatz}
\label{eom_ansa}

The equations of motion \eqref{EOM} on the ansatz (\ref{pheno}, \ref{ans_met}, \ref{ans_met2}, \ref{ans_gau}, \ref{ans_sca}), 
after some simple manipulations, take the form of
\begin{align}
  2 \,C_a^2\, (\beta +\gamma ) (2 \,\beta +2\, \gamma -1) -2\, |V_0|\, r^{-2 \,\gamma +2 \,\delta\,  \kappa +2} + k^2 \,r^{2-2 \,\beta - 2 \,\gamma + 2\, \kappa\, \lambda}  = 0\ , \nonumber \\ \nonumber \\
   C_a^2 \,\kappa \,(2\, \beta +2 \,\gamma - 1) - \alpha\, \left(B^2\, r^{2\, \alpha\,  \kappa }-Q^2 \,r^{-2\, \alpha  \,\kappa }\right) r^{-2 \,(2\, \beta +\gamma -1)} \qquad \qquad \qquad \nonumber \\ \nonumber 
 \qquad \qquad \qquad \qquad + \frac{1}{2}\, \delta \, |V_0| \,r^{-2 \,\gamma +2 \,\delta  \,  \kappa +2} 
  - \frac{1}{2} \,\lambda \,k^2 \,r^{2-2\,\beta-2\,\gamma+2\,\kappa\,\lambda}= 0 \ ,\nonumber \\ 
  C_a^2\, (\beta ^2 +2 \,\beta\,  \gamma - \kappa ^2) + \left(Q^2\, r^{-2 \,\alpha\,  \kappa }+B^2 \,r^{2 \,\alpha  \,\kappa }\right) \,r^{-2\, (2\, \beta +\gamma -1)} \qquad \qquad \qquad  \nonumber \\ \nonumber 
 \qquad \qquad \qquad \qquad -\frac{1}{2}\, |V_0| \,r^{-2 \,\gamma +2 \,\delta\,  \kappa +2} +\frac{1}{2}\, k^2\, r^{2-2\,\beta-2\gamma+2\,\kappa\,\lambda}= 0\ ,\\ \nonumber \\
 (\beta -1)\, \beta +\kappa ^2 = 0 \ .\label{eq_ansatz}
\end{align}

\section{Wider exploration}

In this Appendix we examine both the solutions that have not been examined in the main text and repeat the analysis 
of the two already illustrated as a way of checking the results and bridging the approaches of \cite{Kundu:2012jn} and \cite{Gouteraux:2014hca}.
We work directly adopting the $\theta$, $z$ parametrization, namely the solutions are obtained by solving the equations 
of motion for the action \eqref{modello}, using the ansatz:
\begin{eqnarray}
&\qquad \qquad ds^2=-\tilde r^{\theta-2z}dt^2+L^2\tilde r^{\theta-2}d\tilde r^2+\tilde r^{\theta-2}(dx^2+dy^2) \ ,\nonumber \\
&f(\phi)=e^{2\alpha \phi} \, \qquad V(\phi)=-\left|V_0\right| e^{2\delta \phi} \ , \qquad Y(\phi)=e^{2\lambda \phi} \ , \\
&\qquad \qquad  A = \tilde Q \tilde r^{\zeta - z} dt - B\,  y\, dx\ , \qquad \phi = \tilde \kappa \ln(\tilde r)\ , \qquad \psi^I = k x^I\ , \nonumber
\end{eqnarray}
where $\zeta = 2(1 - \alpha \tilde \kappa)$ is the IR anomalous dimension of the charge density 
and is obtained by solving the Maxwell equation.

The NEC conditions \eqref{nec} (which include $\tilde\kappa>0$) and the consistency condition coming 
from the positivity of the specific heat constitute general requirements valid for all the classes here presented;
they fix the allowed domain to coincide with the colored region in the right panel of Fig \ref{domino} (either color).
Throughout all the cases listed here we have 
\begin{equation}
 \tilde\kappa^2 = (\theta-2)(\theta+2-2z)\ .
\end{equation}

We exclude from the exploration all the solutions where the momentum-dissipation scalar correspond to 
irrelevant operators. As discussed in Subsection \ref{other}, in such cases the transport 
coefficient gets corrected at the quadratic in $k$, this being the same order that we neglect
in the background.

\subsection{No charge density, marginally relevant magnetic field}

Consider $Q=0$ and non-trivial $B$, namely the ``electro-magnetic'' dual configuration 
of the charged solution in the main text. The marginal relevance of $B$, the momentum-dissipation scalars 
and the dilaton leads to
\begin{equation}\label{marre}
 2 \tilde\kappa \alpha = \theta - 4\ , \qquad 
 \tilde\kappa \lambda + 1 = 0\ , \qquad 
 2 \tilde\kappa \delta + \theta = 0\ . 
\end{equation}
Solving the equation we obtain
\begin{equation}
 B^2 = \frac{k^2 (\theta -2 z)+2 \left|V_0\right| (z-1)}{4\,(\theta -z-1)}\ , \qquad 
 L^2 = \frac{2 (\theta -z-1) (\theta -z-2)}{2 \left|V_0\right|-k^2} \ .
\end{equation}
The validity domain is given by 
\begin{equation}
 \left|V_0\right|>\frac{k^2 (2 z-\theta )}{2 (z-1)}
\end{equation}
intersected with the following set of alternative conditions
\begin{align}
 &2 z>\theta +2\ \ \land\ \ 0<\theta <2 \qquad \text{or}\\
 &\theta >2 \land z<0 \qquad \text{or}\\
 &z>1\land \theta \leq 0\ .
\end{align}

\subsection{Marginally relevant charge, magnetic field and momentum-dissipation scalars}

The system encoding the marginality conditions for all the operators reads 
\begin{align}
 2 \delta  \tilde\kappa +\theta &=2 \tilde\kappa  \lambda +2\\
 \tilde\kappa  (\gamma +\delta )+4&=2 \theta\\
 2 \delta  \tilde\kappa +\theta &=0\\
 \tilde\kappa  (\alpha +\delta )+\theta &=2\ .
\end{align}
This system however is impossible.

\subsection{Irrelevant charge density, marginally relevant magnetic field}

The marginal relevance of $B$, the momentum-dissipation scalars and the dilaton 
leads to \eqref{marre}. Solving the EOM one gets 
\begin{equation}
 \begin{split}
 &\theta = 4 + 2 \alpha \tilde\kappa\ , \qquad 
 B^2 = \frac{k^2 (\theta -2 z)+2 \left|V_0\right| (z-1)}{4\,(\theta -z-1)}\ , \\ 
 & \qquad \qquad L^2 = \frac{2 (\theta -z-1) (\theta -z-2)}{2 \left|V_0\right|-k^2}
 \end{split}
\end{equation}
The validity domain (including the condition enforcing the irrelevance of the charge density)
are given by 
\begin{equation}
 \left|V_0\right|>\frac{k^2 (2 z-\theta )}{2 (z-1)}\ ,
\end{equation}
intersected with 
\begin{equation}
 2<\theta <4\ \ \land\ \  z<0 \qquad \text{or} \qquad \theta >4\land z<0\ .
\end{equation}

\subsection{Marginally relevant charge density, irrelevant magnetic field}

This corresponds to the charged solutions already considered in the main text.
Marginal relevance for the charge density, the momentum-dissipation scalar and the dilaton yields
\begin{equation}
 \tilde\kappa \lambda +1 = 0\ , \qquad 
 2\tilde\kappa \delta +\theta = 0 \ , \qquad 
 2\tilde\kappa \alpha = 4 - \theta\ .
\end{equation}
The irrelevance of the magnetic field corresponds to
\begin{equation}
 \theta < 4\ .
\end{equation}
Solving the EOM's we further get
\begin{equation}
 \tilde Q^2 =\frac{k^2 (2 z-\theta )-2 \left|V_0\right| (z-1)}{2\left(2 \left|V_0\right|-k^2\right) (\theta -z-2)}\ , \qquad 
 L^2 = \frac{2 (\theta -z-1) (\theta -z-2)}{2 \left|V_0\right| - k^2}\ ;
\end{equation}
the quantities on the RHS's must therefore be positive.
All in all, the validity domain of the solution is
\begin{equation}
 2<\theta <4\ \land\ z<0 \qquad \text{and} \qquad \left|V_0\right| > \frac{k^2 (2 z-\theta )}{2 (z-1)}\ .
\end{equation}
We agree with \eqref{xiregime} and \eqref{chargedregime3} where
$\xi = k^2/\left|V_0\right|$.

\subsection{Irrelevant charge density and magnetic field}

This corresponds to the neutral solution already considered in the main text.
Asking for marginal of the momentum-dissipation scalars and the dilaton leads to
\begin{equation}
 \tilde\kappa \lambda +1 = 0\ , \qquad 
 2\tilde\kappa \delta +\theta = 0 \ .
\end{equation}
Irrelevance of the charge density and the magnetic field requires
\begin{equation}\label{irre}
 \theta - 4 + 2\alpha \tilde\kappa < 0\ , \qquad
 \theta - 4 - 2\alpha \tilde\kappa < 0 \ .
\end{equation}
which corresponds to \eqref{neutralboh}.
Solving the EOM's we get
\begin{equation}
 \left|V_0\right|  = \frac{k^2 (2 z-\theta )}{2 (z-1)}\ , \qquad 
 L^2 = \frac{2 (z-1) (-\theta +z+2)}{k^2}\ .
\end{equation}
Eventually, the validity domain is
\begin{equation}
 2<\theta <4 \qquad \text{and} \qquad  z<0\ ,
\end{equation}
to be considered along with \eqref{irre}.

\section{Temperature scaling of the Seebeck coefficient}

The definition of the Seebeck coefficient $s$ is
\begin{equation}
s\,=\,\frac{\alpha_{xx}}{\sigma_{xx}}\ .
\end{equation}
From the transport coefficients (\ref{alpha}, \ref{resistivity}) we can extract the generic form of the Seebeck:
\begin{equation}\label{seebach}
s\,=\,\frac{q \,S}{b^2\, f\, k^2\, Y+B^2\, f^2+q^2}\,\Big|_h\ .
\end{equation}
Considering the $T$ scalings of the various quantities appearing in \eqref{seebach}, we obtain
\begin{align}
s\,&=\,\frac{q (\zeta\, T)^{\frac{2\, \beta }{2 \,\gamma -1}}}{B^2 \,(\zeta\, T)^{\frac{4\, \alpha \, \kappa }{2\, \gamma -1}}+k^2 \,(\zeta\, T)^{\frac{2 \kappa \, (\alpha +\lambda
   )+ 2\beta}{2 \,\gamma -1}}+q^2} \ . \label{morecoeffscalings}
\end{align}

\subsection{Charged solutions}
Expanding up to linear order in $B$ we get
\begin{align}
s\,&=\,\frac{q (\zeta\, T)^{\frac{2\, \beta }{2 \,\gamma -1}}}{k^2 \,(\zeta\, T)^{2\frac{\kappa (\alpha +\lambda) + \beta}{2 \,\gamma -1}}+q^2} + {\cal O}(B^2)\ . 
\end{align}
Considering relations \eqref{soluzzo} and the domain of definition of the charged solutions \eqref{chargeregime}, at low $T$ we have
\begin{align}
s\,&=\,\frac{q}{k^2 +q^2} (\zeta\, T)^{\frac{2\, \beta }{2 \,\gamma -1}} + {\cal O}(B^2)\ . 
\end{align}

\subsection{Neutral solutions}

Expanding up to linear order in both $B$ and $q$ we obtain
\begin{align}
s\,& = \,\frac{q }{k^2}\ (\zeta\, T)^{-\frac{2 \kappa \, (\alpha +\lambda)}{2 \,\gamma -1}} + {\cal O}(B^2,q^2) \ . 
\end{align}

\end{document}